\documentclass{article}

\usepackage[final]{neurips_2022_ml4ps}

\usepackage[utf8]{inputenc} 
\usepackage[T1]{fontenc}    
\usepackage{url}            
\usepackage{booktabs}       
\usepackage{amsfonts}       
\usepackage{nicefrac}       
\usepackage{microtype}      
\usepackage{graphicx}
\usepackage{amsmath}
\usepackage{graphicx}
\usepackage[colorlinks=true, allcolors=blue]{hyperref}
\usepackage[position=top]{subfig}
\usepackage[parfill]{parskip}
\usepackage{array}
\usepackage{booktabs}
\usepackage[dvipsnames]{xcolor}
\usepackage{float}
\usepackage{multicol}

\title{Simulation-based inference of the 2D ex-situ stellar mass fraction distribution of galaxies using variational autoencoders}

\author{%
  Eirini Angeloudi \\
  Institute of Astrophysics of the Canary Islands\\
  E-38205 La Laguna, Tenerife, Spain\\
  \texttt{eirini@iac.es} \\
   \And
   Marc Huertas-Company \\
    Institute of Astrophysics of the Canary Islands\\
    E-38205 La Laguna, Tenerife, Spain\\
    \texttt{mhuertas@iac.es } \\
   \AND
   Jesús Falcón-Barroso \\
    Institute of Astrophysics of the Canary Islands\\
    E-38205 La Laguna, Tenerife, Spain\\
    \texttt{jfalcon@iac.es} \\
    \And
   Regina Sarmiento \\
    Institute of Astrophysics of the Canary Islands\\
    E-38205 La Laguna, Tenerife, Spain\\
    \texttt{regina.sarmiento@iac.es} \\
   \AND
   Daniel Walo-Martín \\
    Institute of Astrophysics of the Canary Islands\\
    E-38205 La Laguna, Tenerife, Spain\\
    \texttt{daniel.walo.martin@iac.es} \\
    \And
   Annalisa Pillepich \\
    Max Planck Institute for Astronomy\\
    Königstuhl 17, 69117 Heidelberg, Germany\\
    \texttt{pillepich@mpia.de} \\
   \AND
   Jesús Vega Ferrero \\
    Institute of Astrophysics of the Canary Islands\\
    E-38205 La Laguna, Tenerife, Spain\\
    \texttt{jesus.vega@iac.es} \\
}

\begin{document}

\maketitle

\begin{abstract}
 Galaxies grow through star formation (in-situ) and accretion (ex-situ) of other galaxies. Reconstructing the relative contribution of these two growth channels is crucial for constraining the processes of galaxy formation in a cosmological context. In this on-going work, we utilize a conditional variational autoencoder along with a normalizing flow - trained on a state-of-the-art cosmological simulation -  in an attempt to infer the posterior distribution of the 2D ex-situ stellar mass distribution of galaxies solely from observable two-dimensional maps of their stellar mass, kinematics, age and metallicity. Such maps are typically obtained from large Integral Field Unit Surveys such as MaNGA. We find that the average posterior provides an estimate of the resolved accretion histories of galaxies with a mean $\sim 10\%$ error per pixel. We show that the use of a normalizing flow to conditionally sample the latent space results in a smaller reconstruction error. Due to the probabilistic nature of our architecture, the uncertainty of our predictions can also be quantified. To our knowledge, this is the first attempt to infer the 2D ex-situ fraction maps from observable maps.
\end{abstract}

\section{Introduction}
The standard cosmological model, known as $\Lambda$CDM, suggests that galaxies evolve by successively merging into more massive structures. The mass accreted through this merging/interaction with other galaxies is often referred to as the ex-situ mass of the galaxy. Contrarily, mass created through internal star formation sustained by cosmic gas inflows is referred to as the galaxy's in-situ stellar mass.

Unveiling how the ex-situ mass of a galaxy is distributed within a galaxy can provide an unprecedented glimpse on its cosmic evolution \citep{evolution}. However, the mapping between observable properties - e.g. kinematics or chemical composition of stars - and the accretion history of a galaxy is unknown and intractable. Consequently, there does not exist a clear path in literature that allows the inference of the distribution of the accreted stellar mass of a galaxy. To this end, we propose to resort to simulation-based inference \citep{sib}, using state-of-the art hydrodynamic cosmological simulations. 

Our approach to model the posterior distribution of resolved ex-situ mass in a galaxy consists of a conditional variational autoencoder (cVAE)  that uses as input the 2D distribution maps of ex-situ mass fraction of galaxies - accessible in the simulations - and as a condition the observable 2D maps of their stellar mass, kinematics, age and metallicity. The cVAE is trained to reconstruct the ex-situ maps while learning their multivariate latent distribution. By sampling from this latent space and along with the condition maps, we are able to generate samples from the posterior distribution of ex-situ maps conditioned on the observables. To further enhance the reconstruction precision of the ex-situ maps, we utilize a normalizing flow that enables a finer sampling of the latent space and relaxes the Gaussian prior on the cVAE latent space.

\section{Data}\label{data}

In this first work, our simulator to perform the inference consists of a unique cosmological hydrodynamical simulation TNG100 \citep{tng1, Pillepich_2017, Springel_2017, Nelson_2017, Naiman_2018, Marinacci_2018}. We select galaxies with stellar mass $> {10}^{10}\ {M}_{\odot }$ at a redshift $z = 0$ and $z = 0.1$. Galaxies are forward modeled to produce observables which are used as conditions for our cVAE. We namely produce two-dimensional projections of stellar particle properties (mass, velocity, velocity dispersion, age, metallicity) along a given line-of-sight. For each galaxy, the field of view (FoV) is selected as four times its half-mass radius and the resulting maps have a fixed size of 128 x 128 pixels. 

In order to construct the ground-truth that we wish to infer, namely the ex-situ mass distribution maps, we use the catalog created by \cite{tng_exsitu1, tng_exsitu2} that classifies all stellar particles either as in-situ or ex-situ by following the merger tree of their host galaxy. From there, it is straightforward to produce 2D maps, where each pixel holds the binned average of ex-situ mass fraction of all stellar particles present in that projection. 

Our dataset is highly unbalanced, as galaxies with a mean low ex-situ fraction prevail and there exist relatively few galaxies with a mean high ex-situ mass fraction $(> 0.8)$. To address this issue as well as to increase our training sample size, we create multiple realisations for each galaxy by using different 2D projections of their properties. The number of projections is decided by taking into account the balancing of the dataset. More specifically, a higher number of projections is produced for galaxies with a high ex-situ fraction to account for their rareness and a lower number of projections is produced for galaxies with a low ex-situ fraction. Each line-of-sight is considered as an independent observation.

Through this process, we produce a dataset of 24248 objects that are then randomly split into a training (75\%), a validation (10\%) and a test subset (15\%). Extra caution is taken across splitting to ensure that all alignments of one galaxy are gathered in the same dataset type, so that to avoid leaks of information when testing the prediction results of our model.

\section{Model}\label{model}

We construct a conditional variational autoencoder (cVAE)~\citep{cvae}, to estimate the posterior distribution of the 2D ex-situ stellar mass fraction ($\epsilon$) conditioned on the 2D maps of the stellar properties ($\theta$):
$P(\epsilon|\theta)=\int p(\epsilon|z,\theta).p(z|\theta)dz$, where $p(\epsilon|z,\theta)$ is the generator network and $z$ is the latent space variable of the cVAE which is modeled as a multivariate Gaussian distribution.

\begin{figure}[H]
    \centering
    \includegraphics[width=\linewidth]{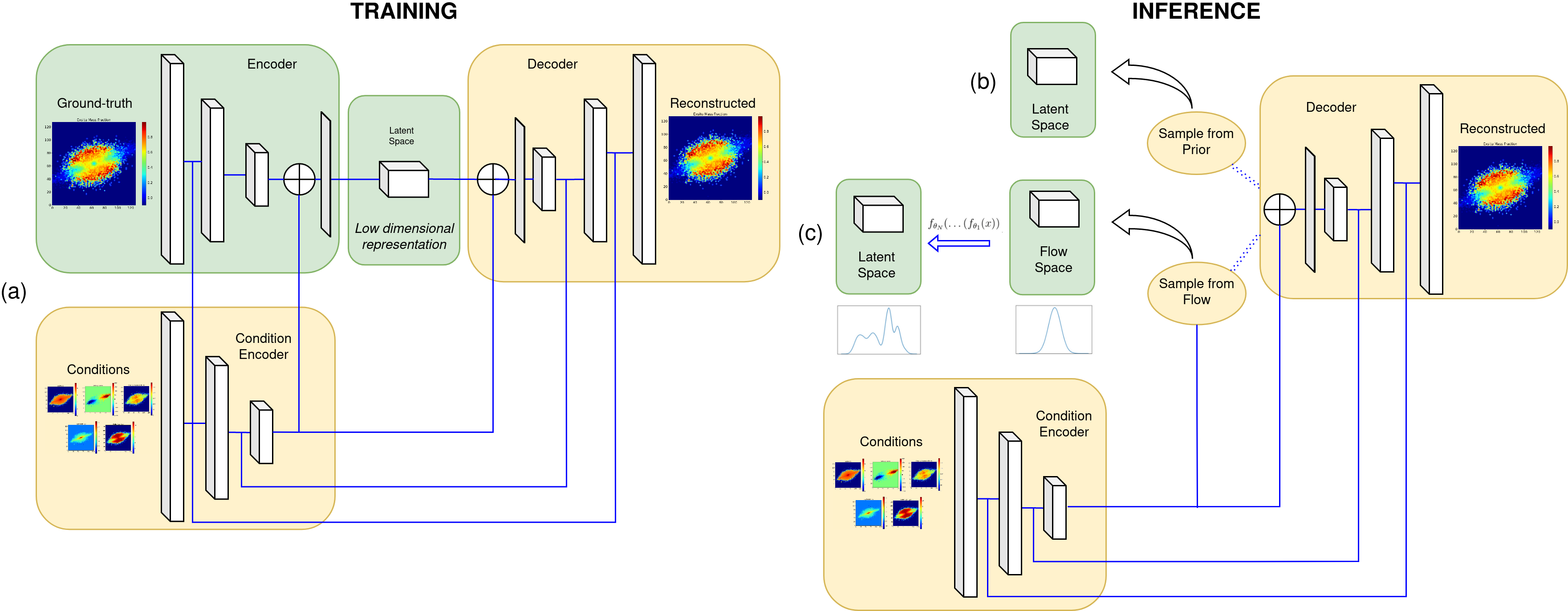}
    \caption{The architecture of the model consisting of a cVAE and optionally a normalizing flow. (a) The architecture of the cVAE during training. This model can also be used for inference when the ground truth is supplied as input to the encoder to acquire a lower bound of the reconstruction error the model is capable of. (b) The model during inference using only the trained cVAE. As the encoder part is removed, we sample from the prior of the latent space and along with the conditions, the ex-situ map is predicted from the decoder. (c) The model during inference when a trained normalizing flow on the latent space is included. We sample conditionally from the flow latent space for a vector compliant with the condition vector and the ex-situ map is predicted from the decoder while the observable maps also contribute to all convolution layers.}
    \label{fig:architecture}
\end{figure}

Our cVAE consists of three CNNs as shown in Figure \ref{fig:architecture}(a): the encoder, the condition encoder and the decoder. Since our conditions are not scalar values, but instead a set of 5 observable 2D maps, we encode them using a CNN. Notably, we add skip connections between the conditions and the encoder/decoder in all convolutional layers to improve the reconstruction as typically done with U-nets. A similar architecture has also been utilized in \cite{hyphy}. The cVAE is trained with a standard evidence lower bound (ELBO) loss function which includes the condition term $\theta = q_{\Psi}(X)$ corresponding to the summary statistics extracted from the condition encoder:
\begin{equation}
    \mathcal{L} = - E_{z\sim p_{\Theta}(z | \epsilon, \theta)}[\log{p_{\Phi}(\epsilon | z, \theta)}] + \beta E_{z\sim p_{\Theta}(z | \epsilon, \theta)}[\log{p_{\Theta}(z | \epsilon, \theta)} - \log{\mathcal{N}(0, 1)}], 
\end{equation}

where $X$ are the condition maps, $\Theta$, $\Phi$ and $\Psi$ are the trainable parameters of the encoder, decoder and condition encoder respectively and $\beta$ is the weight enforced on the Kullback–Leibler Divergence term. 

During the training of the cVAE, the three CNN networks are updated simultaneously. Once the cVAE is trained, we isolate the encoder and the condition encoder parts of the network and reuse them on the whole training set to generate samples of the latent space along with their encoded condition vectors. This new dataset is then utilized as the training set of the conditional normalizing flow, based on autoregressive algorithms \citep{Germain, Papamakarios}.

During inference, we propose two different model architectures. In the first one, the decoder is simply fed with samples from the prior and encoded conditions (Figure \ref{fig:architecture} (b)). In the second approach, we utilize the trained normalizing flow, conditioned on the encoded observables, to sample from the latent space (Figure \ref{fig:architecture} (c)). By adding a normalizing flow to our architecture, we expect to reduce the weight of the Kullback–Leibler Divergence term during the training of the cVAE and hence obtain a more accurate reconstruction.  Both approaches produce samples of the posterior distribution of ex-situ stellar mass that we compare in the following section.

\section{Results} \label{results}
\begin{figure}[H]
    \centering
    \includegraphics[width=0.78\textwidth]{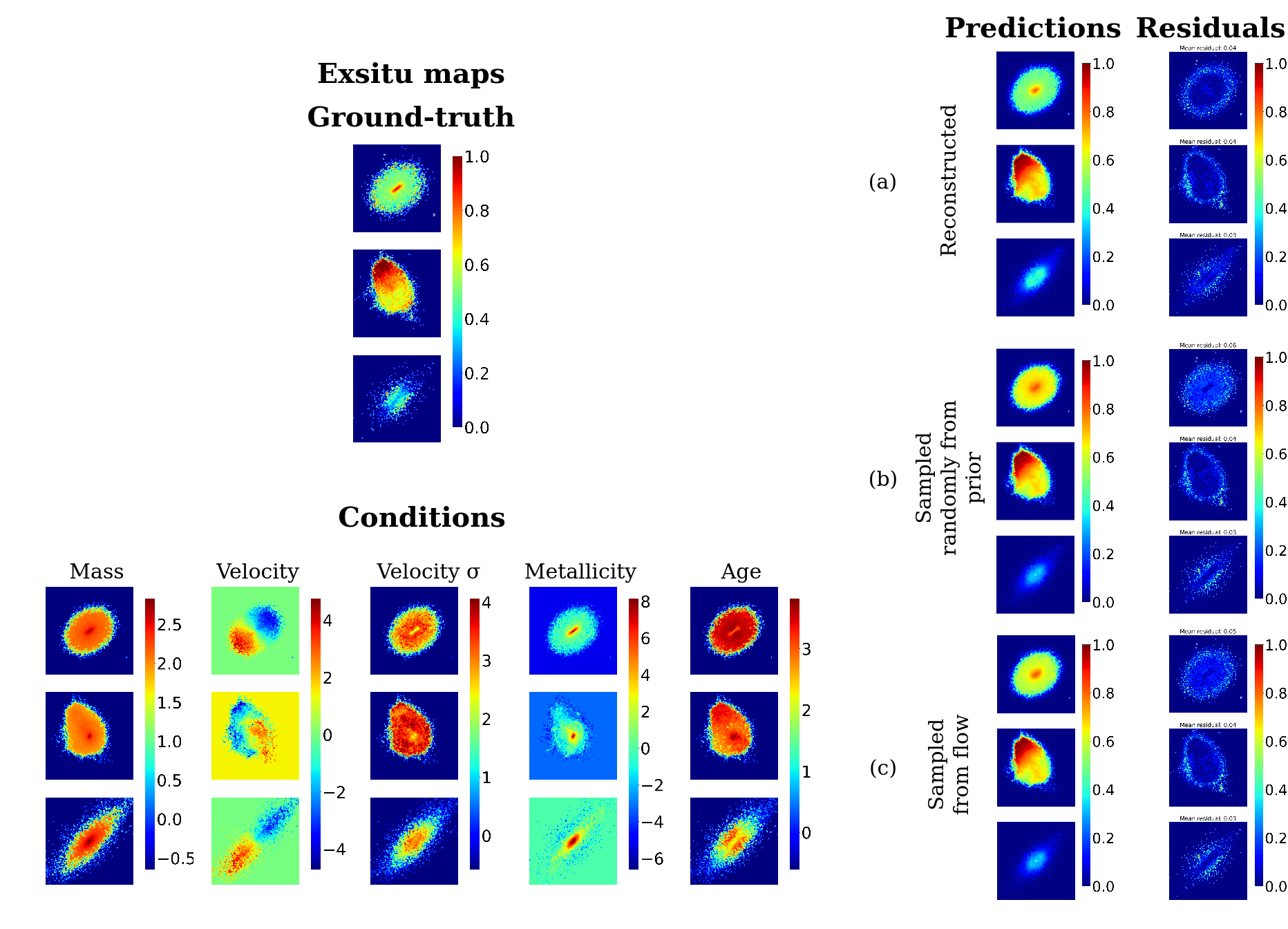}    
    \caption{The results of the Flow-VAE model predictions when evaluated on 3 galaxies from the test set with the ground-truth ex-situ maps and using the observable condition maps on the left. On the right, we show the posterior average calculated on 100 reconstructions of the decoder when supplied along with the conditions with (a) the latent space vectors the encoder outputs on the ground-truth (reconstruction) (b) latent space vectors sampled randomly from the prior (c) latent space vectors sampled from the normalizing flow mapped from the conditions.}
    \label{fig:results}
\end{figure}

We train a variety of models for 200 epochs with different latent space sizes and weights of the Kullback–Leibler divergence term. To optimize the execution time, we train our models on a NVIDIA Tesla P100 GPU. In figure \ref{fig:results}, we illustrate a set of average predicted posteriors on 3 sample galaxies from the test dataset from a model trained with $\beta = 7e-5$ and latent size set to 10, which was among the models that achieved the best results. 

We first evaluate the reconstruction accuracy of the model when applied on unseen data. This serves only as a lower bound on the accuracy we can achieve, as this is the unrealistic case where the actual ground truth is also provided to the model along with the conditions. The results of the two proposed inference architectures are also illustrated: sampling from the prior and sampling through the trained normalizing flow. It is apparent that both scenarios provide predictions of the ex-situ mass fraction distribution that are very close to the ground-truth.

\begin{figure}[h]
    \centering
    \subfloat[]{\includegraphics[width=0.45\textwidth]{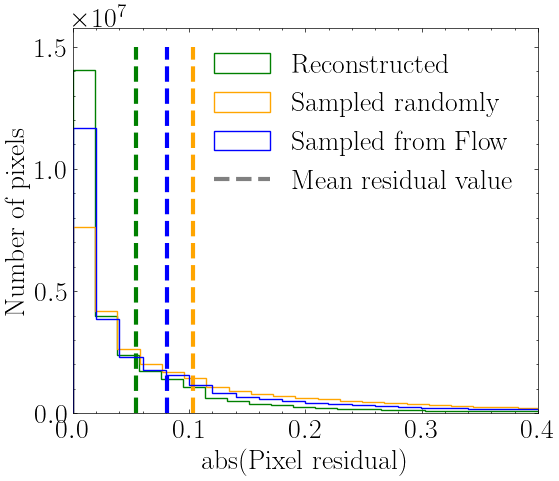}}
    \subfloat[]{\includegraphics[width=0.45\textwidth]{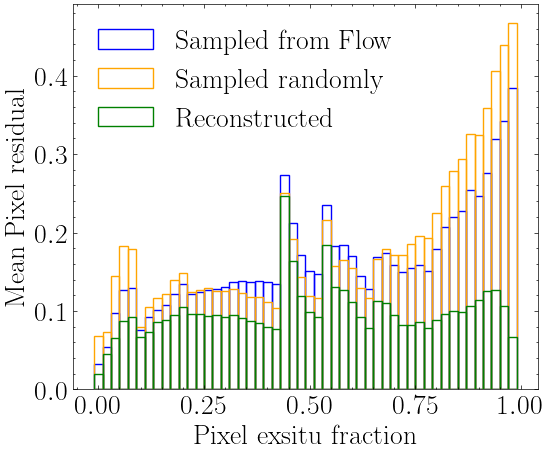}} 
    \caption{Collective results for the model accuracy when the residuals are calculated on the whole test set. (a) The distribution of the pixel residuals when reconstructing the images using the ground-truth (reconstructed), sampling from the flow and sampling randomly from the prior. The mean residual values on the three scenarios are demonstrated as vertical lines colored correspondingly. (b) The correlation of the pixel residuals with the ex-situ fraction of the pixels.}
    \label{fig:resultsall}
\end{figure}

It is worth noting that the extra complexity added by including a normalizing flow in our architecture seems to increase the precision even further. This can be demonstrated in Figure \ref{fig:resultsall} (a), where the distribution of the absolute value of the residuals of all pixels from the test set is shown collectively for the three different scenarios. The distribution of residuals is mostly concentrated around zero, with a higher peak on reconstruction closely followed by the inference through flow sampling. The mean residual values of all pixels are also demonstrated, further validating that the overall prediction power when sampling from the flow is substantially more accurate that when sampling randomly from the prior.

Additionally, we illustrate how residuals are correlated with the pixel ex-situ fraction in Figure \ref{fig:resultsall} (b). It is apparent that reconstruction always provides the lowest residuals, as it was anticipated. Furthermore, we observe that higher ex-situ fraction pixels generally suffer from higher residuals, which can be attributed to their rareness. It is important to stress here that only galaxy pixels are taken into account for the calculation of the residual statistics, to avoid overstating the precision by including background pixels that generally have a very low residual value. 

\section{Future Development} \label{future}

As this work is still in progress, a number of improvements are scheduled for the near future. We plan to apply our model to other cosmological simulations (e.g. EAGLE \citep{eagle}) and train across simulations so that inference is independent of the different sub-grid recipes of each simulation. This crucial step will allow us to apply the model on mock and hopefully actual observation data, enabling the first observational constraint on the accretion histories of galaxies. Additionally, we need to better understand whether the standard deviation of the different realisations can serve as a meaningful measure for capturing the uncertainty of the model. To do so, we should check the coverage probabilities of the per pixel posterior estimates. 

\section{Conclusion}
This on-going work is the first attempt to predict ex-situ fractions of galaxies on a 2D projection and solely from observable maps. By combining simulation data with a variational autoencoder  and a normalizing flow, we are able to predict the posterior distribution of the ex-situ fraction maps with a mean $\sim10\%$ error per pixel. These promising preliminary results suggest that this methodology can serve as a new powerful path to acquire a novel view of the stellar assembly of statistical samples of galaxies from observational surveys.

\section{Impact Statement}
Understanding the origin of stellar mass present in a galaxy can provide a crucial insight on its cosmic evolution. An obvious impediment to study this is that we can only observe galaxies at a specific redshift and the path to correlate observable properties with the accretion history of a galaxy is still unclear. Through this work, we propose the utilization of cosmological simulations, where the evolution history galaxies is traceable, along with machine learning techniques to infer the 2D ex-situ stellar
mass fraction distribution of galaxies. We demonstrate that this experiment works when using a unique cosmological simulation. Our end-goal is to couple multiple cosmological simulations in order to create a robust model that can infer the 2D ex-situ stellar mass fraction distribution of galaxy independently of the dataset that it originates. This will allow for application on actual observational data and additionally the uncertainty of the prediction will also be measured. More importantly, this work can serve as an indicator that machine learning coupled with a proper combination of more than one cosmological simulations can serve as a powerful tool to bridge the gap between observations and cosmological simulations, which can have a substantial impact on a variety of research lines.

\clearpage

\bibliographystyle{mnras}
\bibliography{example} 

\begin{thebibliography}{}
\makeatletter
\relax
\def\mn@urlcharsother{\let\do\@makeother \do\$\do\&\do\#\do\^\do\_\do\%\do\~}
\def\mn@doi{\begingroup\mn@urlcharsother \@ifnextchar [ {\mn@doi@}
  {\mn@doi@[]}}
\def\mn@doi@[#1]#2{\def\@tempa{#1}\ifx\@tempa\@empty \href
  {http://dx.doi.org/#2} {doi:#2}\else \href {http://dx.doi.org/#2} {#1}\fi
  \endgroup}
\def\mn@eprint#1#2{\mn@eprint@#1:#2::\@nil}
\def\mn@eprint@arXiv#1{\href {http://arxiv.org/abs/#1} {{\tt arXiv:#1}}}
\def\mn@eprint@dblp#1{\href {http://dblp.uni-trier.de/rec/bibtex/#1.xml}
  {dblp:#1}}
\def\mn@eprint@#1:#2:#3:#4\@nil{\def\@tempa {#1}\def\@tempb {#2}\def\@tempc
  {#3}\ifx \@tempc \@empty \let \@tempc \@tempb \let \@tempb \@tempa \fi \ifx
  \@tempb \@empty \def\@tempb {arXiv}\fi \@ifundefined
  {mn@eprint@\@tempb}{\@tempb:\@tempc}{\expandafter \expandafter \csname
  mn@eprint@\@tempb\endcsname \expandafter{\@tempc}}}

\bibitem[\protect\citeauthoryear{{Cooper} et~al.,}{{Cooper}
  et~al.}{2010}]{evolution}
{Cooper} A.~P.,  et~al., 2010, \mn@doi [Monthly Notices of the Royal
  Astronomical Society] {10.1111/j.1365-2966.2010.16740.x}, \href
  {https://ui.adsabs.harvard.edu/abs/2010MNRAS.406..744C} {406, 744}

\bibitem[\protect\citeauthoryear{Cranmer, Brehmer  \& Louppe}{Cranmer
  et~al.}{2020}]{sib}
Cranmer K.,  Brehmer J.,   Louppe G.,  2020, \mn@doi [Proceedings of the
  National Academy of Sciences] {10.1073/pnas.1912789117}, 117, 30055

\bibitem[\protect\citeauthoryear{Germain, Gregor, Murray  \&
  Larochelle}{Germain et~al.}{2015}]{Germain}
Germain M.,  Gregor K.,  Murray I.,   Larochelle H.,  2015, MADE: Masked
  Autoencoder for Distribution Estimation, \mn@doi{10.48550/ARXIV.1502.03509},
  \url {https://arxiv.org/abs/1502.03509}

\bibitem[\protect\citeauthoryear{Horowitz, Dornfest, Lukić  \&
  Harrington}{Horowitz et~al.}{2021}]{hyphy}
Horowitz B.,  Dornfest M.,  Lukić Z.,   Harrington P.,  2021, HyPhy: Deep
  Generative Conditional Posterior Mapping of Hydrodynamical Physics,
  \mn@doi{10.48550/ARXIV.2106.12675}, \url {https://arxiv.org/abs/2106.12675}

\bibitem[\protect\citeauthoryear{Marinacci et~al.,}{Marinacci
  et~al.}{2018}]{Marinacci_2018}
Marinacci F.,  et~al., 2018, \mn@doi [Monthly Notices of the Royal Astronomical
  Society] {10.1093/mnras/sty2206}

\bibitem[\protect\citeauthoryear{Naiman et~al.,}{Naiman
  et~al.}{2018}]{Naiman_2018}
Naiman J.~P.,  et~al., 2018, \mn@doi [Monthly Notices of the Royal Astronomical
  Society] {10.1093/mnras/sty618}, 477, 1206

\bibitem[\protect\citeauthoryear{Nelson et~al.,}{Nelson
  et~al.}{2017}]{Nelson_2017}
Nelson D.,  et~al., 2017, \mn@doi [Monthly Notices of the Royal Astronomical
  Society] {10.1093/mnras/stx3040}, 475, 624

\bibitem[\protect\citeauthoryear{Nelson et~al.,}{Nelson et~al.}{2018}]{tng1}
Nelson D.,  et~al., 2018, The IllustrisTNG Simulations: Public Data Release,
  \mn@doi{10.48550/ARXIV.1812.05609}, \url {https://arxiv.org/abs/1812.05609}

\bibitem[\protect\citeauthoryear{Papamakarios, Pavlakou  \&
  Murray}{Papamakarios et~al.}{2017}]{Papamakarios}
Papamakarios G.,  Pavlakou T.,   Murray I.,  2017, Masked Autoregressive Flow
  for Density Estimation, \mn@doi{10.48550/ARXIV.1705.07057}, \url
  {https://arxiv.org/abs/1705.07057}

\bibitem[\protect\citeauthoryear{Pillepich et~al.,}{Pillepich
  et~al.}{2017}]{Pillepich_2017}
Pillepich A.,  et~al., 2017, \mn@doi [Monthly Notices of the Royal Astronomical
  Society] {10.1093/mnras/stx3112}, 475, 648

\bibitem[\protect\citeauthoryear{Rodriguez-Gomez et~al.,}{Rodriguez-Gomez
  et~al.}{2015}]{tng_exsitu1}
Rodriguez-Gomez V.,  et~al., 2015, \mn@doi [Monthly Notices of the Royal
  Astronomical Society] {10.1093/mnras/stw456}, 458

\bibitem[\protect\citeauthoryear{Rodriguez-Gomez et~al.,}{Rodriguez-Gomez
  et~al.}{2016}]{tng_exsitu2}
Rodriguez-Gomez V.,  et~al., 2016, \mn@doi [Monthly Notices of the Royal
  Astronomical Society] {10.1093/mnras/stw456}, 458, 2371

\bibitem[\protect\citeauthoryear{{Schaye} et~al.,}{{Schaye}
  et~al.}{2015}]{eagle}
{Schaye} J.,  et~al., 2015, {The EAGLE project: simulating the evolution and
  assembly of galaxies and their environments} (\mn@eprint {arXiv}
  {1407.7040}), \mn@doi{10.1093/mnras/stu2058}

\bibitem[\protect\citeauthoryear{Sohn, Lee  \& Yan}{Sohn et~al.}{2015}]{cvae}
Sohn K.,  Lee H.,   Yan X.,  2015, Learning Structured Output Representation
  using Deep Conditional Generative Models, \url
  {https://proceedings.neurips.cc/paper/2015/file/8d55a249e6baa5c06772297520da2051-Paper.pdf}

\bibitem[\protect\citeauthoryear{Springel et~al.,}{Springel
  et~al.}{2017}]{Springel_2017}
Springel V.,  et~al., 2017, \mn@doi [Monthly Notices of the Royal Astronomical
  Society] {10.1093/mnras/stx3304}, 475, 676

\makeatother
\end{thebibliography}

\clearpage

\end{document}